\documentclass{appolb}
\usepackage{epsfig}

\def\beq{\begin{equation}}
\def\eeq{\end{equation}}
\def\bea{\begin{eqnarray}}
\def\eea{\end{eqnarray}}

\def\bra#1{\langle #1|}
\def\ket#1{| #1\rangle}
\def\roughly#1{\mathrel{\raise.3ex\hbox
{$#1$\kern-.75em\lower1ex\hbox{$\sim$}}}}

\def\Bbar{{\overline B}^0}

\def \thl {{\theta_l}}
\def \thK {{\theta_{K^*}}}
\def \azeL{{A_0^L}}
\def \azeR{{A_0^R}}
\def \apaL{{A_\parallel^L}}
\def \apaR{{A_\parallel^R}}
\def \apeL{{A_\perp^L}}
\def \apeR{{A_\perp^R}}
\def \re{\text{Re}}
\def \im{\text{Im}}
\def \kstar{{K^*}}
\def \eff{{\text{eff}}}
\def \braket#1#2#3{\langle #1|#2| #3\rangle}

\def \av#1{\left\langle #1\right\rangle}
\def \all{A_{0L}}
\def \alr{A_{0R}}
\def \apl{A_{\| L}}
\def \apr{A_{\| R}}
\def \appl{A_{\bot L}}
\def \appr{A_{\bot R}}
\def \al{A_0}
\def \ap{A_{\|}}
\def \app{{A}_{\bot}}
\def \App{{\mathcal A}_{\bot}}
\def \All{{\mathcal A}_{0L}}
\def \Alr{{\mathcal A}_{0R}}
\def \Apl{{\mathcal A}_{\| L}}
\def \Apr{{\mathcal A}_{\| R}}
\def \Appl{{\mathcal A}_{\bot L}}
\def \Appr{{\mathcal A}_{\bot R}}
\def \Al{{\mathcal A}_0}
\def \Ap{{\mathcal A}_{\|}}
\def \App{{\mathcal A}_{\bot}}

\def \nnu{\nonumber}
\def \bra#1{\left\langle #1\right|}
\def \braket#1#2#3{\langle #1|#2| #3\rangle}
\def \ket#1{\left| #1\right\rangle}
\def \dis{\displaystyle}

\def \Aim{\ensuremath{A_{\mathrm{im}}}\xspace}
\def \AT#1{\ensuremath{A_T^{\left(#1\right)}}\xspace}
\def \FL{\ensuremath{F_{\mathrm{L}}}\xspace}
\def \zerocrossing{\ensuremath{q^2_0}\xspace}

\def \fig#1{Fig.~\ref{#1}}
\def \figs#1{Figs.~\ref{#1}}

\def \thetaL{\ensuremath{\theta_l}\xspace}
\def \thetaK{\ensuremath{\theta_\kaon}\xspace}
\def \qsquare{\ensuremath{q^{2}}\xspace}

\def \Bbar{\overline{\kern -0.24em B}}

\def \mhat{\hat{m}_\kstar}
\def \eff{\mathrm{eff}}
\def \heff{H_{\text{eff}}}
\def \kstar{{K^*}}
\def \trans{{T}}
\def \del#1{\Delta^{#1}_{\tilde D_L}}
\def \chargino{\tilde{\chi}^{\pm}}
\def \sh{\hat{s}}
\def \Im{{\text{Im}}}
\def \Re{{\text{Re}}}

\def \cseff{ C_7^{\text{eff}}}
\def \cseffP{ {C_7^{\text{eff}}}^\prime}
\def \ceff{ C_9^{\text{eff}}}
\def \ceffP{ {C_9^{\text{eff}}}^\prime}
\def \ceffstar{C_9^{\text{eff}*}}
\def \cten{ C_{10}}
\def \ctenP{C_{10}^\prime}
\def \qsquare{\ensuremath{q^{2}}\xspace}

\begin{document}
\title{ The exclusive $B\to K^*(\to K\pi)l^+l^-$ decay:\\ CP conserving observables.%
\thanks{Presented by J. Matias at Flavianet Meeting, Kazimierz, 
July 2009, \\
CERN-PH-TH/2009-207, MZ-TH/09-44, IC/HEP/09-14, UAB-FT-673}%
}
\author{U. Egede$^1$, T. Hurth$^{2,3}$, J. Matias$^4$, M. Ramon$^4$, W. 
Reece$^1$
\address{$^1$ Imperial College London, London SW7~2AZ, United Kingdom\\
$^2$ CERN, Dept. of Physics, Theory  Division, CH-1211 Geneva 23, 
Switzerland \\
$^3$ Institute for Physics, Johannes Gutenberg-University, D-55099 Mainz, Germany\\
$^4$ IFAE, Universitat Aut\`onoma de Barcelona, 08193 Bellaterra, 
Barcelona, Spain}
} \maketitle
\begin{abstract}
\noindent We study the K*  polarization states in the exclusive 4-body \textit{B} meson decay 
$B^0 \to K^{*0}(\to K^- \pi^+)l^+ l^-$  in the low dilepton mass region 
working in the framework of QCDF. 
We 
review the construction of the CP conserving transverse and 
transverse/longitudinal 
observables $A_T^2$, $A_T^3$ and $A_T^4$. We focus here, on 
analyzing their 
behaviour at large recoil energy in 
presence of right-handed currents. 
\end{abstract}
\PACS{13.25.Hw,11.30.Hv,12.39.St}

\def \thl {{\theta_l}}
\def \thK {{\theta_{K^*}}}
\def \azeL{{A_0^L}}
\def \azeR{{A_0^R}}
\def \apaL{{A_\parallel^L}}
\def \apaR{{A_\parallel^R}}
\def \apeL{{A_\perp^L}}
\def \apeR{{A_\perp^R}}
\def \re{\text{Re}}
\def \im{\text{Im}}
\def \kstar{{K^*}}
\def \eff{{\text{eff}}}
\def \braket#1#2#3{\langle #1|#2| #3\rangle}

\def \av#1{\left\langle #1\right\rangle}
\def \all{A_{0L}}
\def \alr{A_{0R}}
\def \apl{A_{\| L}}
\def \apr{A_{\| R}}
\def \appl{A_{\bot L}}
\def \appr{A_{\bot R}}
\def \al{A_0}
\def \ap{A_{\|}}
\def \app{{A}_{\bot}}
\def \App{{\mathcal A}_{\bot}}
\def \All{{\mathcal A}_{0L}}
\def \Alr{{\mathcal A}_{0R}}
\def \Apl{{\mathcal A}_{\| L}}
\def \Apr{{\mathcal A}_{\| R}}
\def \Appl{{\mathcal A}_{\bot L}}
\def \Appr{{\mathcal A}_{\bot R}}
\def \Al{{\mathcal A}_0}
\def \Ap{{\mathcal A}_{\|}}
\def \App{{\mathcal A}_{\bot}}

\def \nnu{\nonumber}
\def \bra#1{\left\langle #1\right|}
\def \braket#1#2#3{\langle #1|#2| #3\rangle}
\def \ket#1{\left| #1\right\rangle}
\def \dis{\displaystyle}

\def \Aim{\ensuremath{A_{\mathrm{im}}}\xspace}
\def \AT#1{\ensuremath{A_T^{\left(#1\right)}}\xspace}
\def \FL{\ensuremath{F_{\mathrm{L}}}\xspace}
\def \zerocrossing{\ensuremath{q^2_0}\xspace}

\def \fig#1{Fig.~\ref{#1}}
\def \figs#1{Figs.~\ref{#1}}

\def \thetaL{\ensuremath{\theta_l}\xspace}
\def \thetaK{\ensuremath{\theta_\kaon}\xspace}
\def \qsquare{\ensuremath{q^{2}}\xspace}

\def \be{\begin{equation}}
\def \ee{\end{equation}}
\def \bea{\begin{eqnarray}}
\def \eea{\end{eqnarray}}
\def \ben{\begin{enumerate}}
\def \een{\end{enumerate}}
\def \bit{\begin{itemize}}
\def \eit{\end{itemize}}
\def \Bbar{\overline{\kern -0.24em B}}

\def \mhat{\hat{m}_\kstar}
\def \eff{\mathrm{eff}}
\def \heff{H_{\text{eff}}}
\def \kstar{{K^*}}
\def \trans{{T}}
\def \del#1{\Delta^{#1}_{\tilde D_L}}
\def \chargino{\tilde{\chi}^{\pm}}
\def \sh{\hat{s}}
\def \Im{{\text{Im}}}
\def \Re{{\text{Re}}}

%
\def \cseff{\color{ForestGreen} C_7^{\text{eff}}}
\def \cseffP{\color{RedOrange} {C_7^{\text{eff}}}^\prime}
\def \ceff{\color{ForestGreen} C_9^{\text{eff}}}
\def \ceffP{\color{ForestGreen} {C_9^{\text{eff}}}^\prime}
\def \ceffstar{C_9^{\text{eff}*}}
\def \cten{\color{ForestGreen} C_{10}}
\def \ctenP{C_{10}^\prime}

\date{\today}

\section{Motivation} 

The exclusive decay $B\to K^* l^+l^-$ will play a central role in the 
near future at LHCb and also at Super-LHCb. This channel is particularly 
interesting because it provides information in different ways. It is used as a basis to construct different type of 
observables, such as the forward-backward (FB) asymmetry \cite{fba,fba2}, 
the isospin 
asymmetry \cite{is} and the angular distribution 
observables \cite{km,lm,ehmrr1,ehmrr2,buras,gudrun}. Here, we will focus on 
the 
observables derived from the 4-body decay distribution: $B\to K^*(\to K\pi)l^+l^-$ that provides information on the K* spin amplitudes.

\section{Differential decay distributions, K* Spin Amplitudes and Non Minimal Supersymmetric model}
The starting point is the differential decay distribution of the 
decay ${\bf {\bar B_d}\to {\bar K^{*0}}(\to K\pi)l^+l^-}$. This distribution with the $K^{*0}$ on the mass shell is described by ${ 
s}$ and three angles ${\bf\theta_l}$ (angle between $\mu^-$ and the direction of the outgoing $K^*$ in $\mu\mu$ frame)
, ${\bf\theta_K}$ (angle between $K^-$ and outgoing $K^*$
in $\bar K^*$ frame) 
 and ${\bf\phi}$ (angle between the two planes), 
\begin{equation}
  \label{diff:four-fold}
  \frac{d^4\Gamma}{d{q^2}\,d{ \theta_l}\, d{ \theta_K}\, d{ \phi}} =
  \frac{9}{32 \pi} I({q^2}, {  \theta_l, { \theta_K}, { \phi}}) \sin{ \theta_l}\sin{ \theta_K}\:\nonumber
\end{equation}
where $ I = {    \bf I_1} + {    \bf I_2}\cos 2\theta_l + {    \bf I_3} \sin^2\theta_l\cos 2\phi +
  {    \bf I_4} \sin 2\theta_l \cos\phi + {    \bf I_5} \sin\theta_l\cos\phi + 
   + {    \bf I_6} \cos\theta _l + {    \bf I_7} \sin\theta_l\sin\phi +
{    \bf I_8} \sin 2\theta_l \sin\phi + {    \bf I_9} \sin^2\theta_l\sin 2\phi.$ 
In the massless limit the I's are function of the K* spin amplitudes\cite{km} $A_{\perp L,R}$, $A_{\| L,R}$ and $A_{0 L,R}$,
we have then 6 complex amplitudes, four symmetries (see \cite{ehmrr2}) and 9 $\rm I_i$ parameters, of which 8 are independent.
At this point we can follow two alternatives to construct observables: a) fit the parameters I's, use them as observables, and 
compare 
the predictions with data or b) use the spin amplitudes as the key ingredient to construct a selected group of 
observables. 

The first option (`a')\cite{buras} is experimentally problematic as the resultant fit fails to capture the correlation between the I's induced by the underlying K* spin amplitudes. 
The second option (`b')\cite{ehmrr1} aims at constructing selected 
observables from the K* spin amplitudes that are extracted directly from the experimental fit. Certain criteria are considered: maximal sensitivity to right-handed currents (RH), minimal sensitivity to poorly 
known soft form factors, and good experimental resolution. We 
will always follow option `b'\cite{ehmrr1}. The procedure in this case is the following: choose the combination of spin amplitudes with maximal 
sensitivity to RH 
currents; check if the combination fulfills all symmetries; and finally, analyse the observables and New Physics (NP) impact.
Notice that these combinations of spin amplitudes may be simple functions of the I's (see \cite{ehmrr1}) or highly non-linear 
combinations showing up an interesting sensitivity to NP (see 
\cite{ehmrr2} for an example).

The keypoint is the evaluation of the relevant matrix elements that in naive factorization are 
functions of the form factors $V(q^2)$, $A_{0,1,2,3}(q^2)$ and $T_{1,2,3}(q^2)$. Then the spin amplitudes $A_{\perp,\|,0}$ can be written in 
terms of these form factors and the Wilson coefficients $C_{7}^{\mathrm {eff}}$, $C_7^{\mathrm{eff} 
\prime}$, $C_9^{\mathrm{eff}}$ and $C_{10}$ of an effective Hamiltonian
that includes RH currents via the electromagnetic dipole operator 
$O_7^\prime=(e/16 \pi^2) m_b(\bar s \sigma_{\mu\nu}P_Lb)F^{\mu\nu}$:
{
$$\label{a_perp} 
{\bf A_{\bot L,R}}={\hat N}\lambda^{1/2}\bigg[
({ C^{\rm eff}_{9}}\mp{  C_{10}})\frac{  V(q^2)}{m_B +m_K^*}+\frac{2m_b}{q^2} ({ C^{\rm eff}_{7}} + { C^{\rm eff \prime}_{7}})
{ T_1(q^2)}\bigg] \nonumber
$$}
{
$$\label{a_par} 
{\bf A_{\| L,R}}\!=\!-{\hat N}(m_B^2- m_{K^*}^2)\bigg[({  C^{\rm eff}_{9}}\mp 
{  
C_{10}})
\frac{  A_1 (q^2)}{m_B-m_{K^*}}
+\frac{2 m_b}{q^2} ({ C^{\rm eff}_{7}} - { C^{\rm eff\prime}_{7}}) { T_2(q^2)}\bigg]
\nonumber
$$}
{
\begin{eqnarray}\label{a_long} 
{\bf A_{0L,R}}&\!\!=\!\!&-\frac{\hat N}{2m_{K^*}\sqrt{2q^2}}\nnu \bigg[
({  C^{\rm eff}_{9}}\mp {  C_{10}})\bigg\{(m_B^2-m_{K^*}^2 
-q^2)(m_B+m_{K^*}){  A_1(q^2)}
 \nonumber \\
&& -\lambda \frac{  A_2(q^2)}{m_B +m_{K^*}}\bigg\}
 + {2m_b}({ C^{\rm eff}_{7}} - { C^{\rm eff \prime}_{7}}) \bigg\{
 (m_B^2+3m_{K^*}^2 -q^2){ T_2(q^2)} \nonumber \\
&&  -\frac{\lambda}{m_B^2-m_{K^*}^2} {T_3(q^2)}\bigg\}\bigg], \nonumber
\end{eqnarray}
}
We are left again with two possible choices: either i) use QCD light 
cone sum rules (LCSR) 
to estimate the required form factors adding the $\alpha_s$ corrections 
from QCDF or ii) work consistently in the same framework of QCDF at LO 
and NLO \cite{fba2} and include a reasonable conservative size for the 
possible $\Lambda/m_b$ corrections. The first 
option (`i')\cite{buras} implies neglecting some ${\cal O}(\Lambda/m_b)$ corrections 
to QCDF and assume that the main part of those corrections are inside the soft form 
factors evaluated with QCD LCSR. The second option (`ii')\cite{ehmrr1} allows us to explore the impact that ${\cal O}(\Lambda/m_b)$ corrections have on the observables.

In the limit {  $m_B \to \infty$} and {  $E_K^* \to \infty$} all form factors are related to only two soft form factors 
$\xi_\perp$ and $\xi_\|$ \cite{Charles} and consequently
transversity amplitudes simplify enormously; then observables can be easily constructed in which the soft form factors cancel out completely at LO:
{
$$\label{LEL:tranversity:perp}
{\bf A_{\bot L,R}}= {\hat  N} m_B(1- \sh)\bigg[
({  C^{\rm eff}_{9}}\mp{  C_{10}})+\frac{2\hat{m}_b}{\sh} ({ C^{\rm eff}_{7}} + { C^{\rm eff \prime}_{7}})
\bigg]\xi_{\bot}(E_\kstar), \nonumber
$$
$$\label{LEL:tranversity:par}
{\bf A_{\| L,R}}= -{\hat N} m_B (1-\sh)\bigg[({  C^{\rm eff}_{9}}\mp{  
C_{10}})
+\frac{2\hat{m}_b}{\sh}({ C^{\rm eff}_{7}} - { C^{\rm eff \prime}_{7}})
 \bigg]\xi_{\bot}(E_\kstar), \nonumber
$$
\bea\label{LEL:tranversity:zero}
{\bf A_{0L,R}}= -\frac{{\hat N}m_B }{2 \hat{m}_\kstar \sqrt{2\sh}} 
(1-\sh)^2\bigg[({  C^{\rm eff}_{9}}\mp{  C_{10}}) + 2
\hat{m}_b ({ C^{\rm eff}_{7}} - 
{ C^{\rm eff \prime}_{7}}) \bigg]\xi_{\|}(E_\kstar), \nonumber
\eea
}
where $\sh = q^2/m_B^2$, $\hat{m}_b = m_b/m_B$ and $\hat{m}_\kstar = m_\kstar/m_B$.

Finally, our BSM testing ground model will be a Supersymmetric model with non-minimal flavour violation in the down squark sector that induces RH currents\cite{lm}. We will focus on two scenarios\cite{ehmrr1}:

\begin{itemize}
\item {  \bf Scenario A}: { Large-gluino and positive mass 
insertion scenario:} 
  $m_{\tilde g} = 1 \rm TeV$, $m_{\tilde d} \in
  [200,1000] \rm GeV$,.   
{   Curve (a):} {  $m_{\tilde g}/m_{\tilde d}=2.5$,
  $\left(\delta_{LR}^{d}\right)_{32}=0.016$.}
{   Curve (b):} {  $m_{\tilde g}/m_{\tilde d}=4$,
  $\left(\delta_{LR}^{d}\right)_{32}=0.036$.} 

\item {  \bf Scenario B:} { Low-gluino mass or large squark mass:}  
$m_{\tilde d} = 1 \rm Tev$, $m_{\tilde g} \in
  [200,800] \rm Gev$.
 { Curve (c):} { $m_{\tilde
    g}/m_{\tilde d}=0.7$, $\left(\delta_{LR}^{d}\right)_{32}=-0.004$.} 
 {  Curve (d):} {
  $m_{\tilde g}/m_{\tilde d}=0.6$,
  $\left(\delta_{LR}^{d}\right)_{32}=-0.006$. }
\end{itemize}

We also take: $\mu=M_1=M_2=m_{\tilde uR}=1 \rm TeV$ and $\tan\beta=5$.
All relevant constraints (coming from \textit{B} physics rare decays $\rho$ parameter, Higgs mass, SUSY particle searches, vacuum stability, etc.) have also been checked.

\section{Analysis of observables}

In the framework of QCDF  
at NLO we evaluate the K* spin amplitudes to include $\alpha_s$ 
contributions to form factors, adding 
also  
possible 
$\Lambda/m_b$ corrections according to option `(ii)'\cite{ehmrr1}. 
We are then in the position to construct  observables out of 
these spin 
amplitudes, the so called `transverse and 
transverse/longitudinal asymmetries' \cite{km,lm,ehmrr1}: $A_T^2$, 
$A_T^3$ and $A_T^4$. 
In order  to fully understand the behaviour of these 
observables it is very illuminating to analyze them in the large recoil 
limit using 
the heavy quark and large-$E_{K^*}$  expressions for the spin amplitudes.
This is the main goal of this section.
\newline

The transverse asymmetry $A_T^2$, first proposed in \cite{km}, probes the transverse spin amplitude $A_{\perp,\|}$ in a controlled way. It is defined by \cite{km}:
\begin{equation}
    \label{eq:AT2Def}
  \hspace*{1.2cm}  A_T^{{2}} =\frac{|A_\perp|^2 - |A_\||^2}{|A_\perp|^2 + |A_\||^2} \, \nonumber
  \end{equation}
This observable has a particularly simple form if one uses the heavy quark and large-$E_{K^*}$ limit for 
the transverse amplitudes:
\begin{equation}
    \label{eq:AT2LEET}
   A_{T}^{{2}} \,
{\sim}\, { 4 { C^{\mathrm{eff} \prime}_{7}} \frac{m_b M_B}{q^2} 
\frac{\Delta_{-}+\Delta_{+}^*}{2 C_{10}^2+|\Delta_{-}|^2+|\Delta_{+}|^2}}  \, \nonumber
  \end{equation}
where $\Delta_{\pm}={\cal C}_9^{\rm eff}+ 2 \frac{m_b M_B}{s}
 ({  C^{{\rm eff}}_{7}} \pm  { C^{\rm {eff} \prime}_{7}})$.
It is then clear that in this observable $\xi_{\perp}(0)$ form factor 
dependence cancels 
at LO and the sensitivity to ${C^{{\rm eff} \prime}_{7}}$ is maximal.

We restrict our analysis to the low-dilepton mass region $1 \leq q^2 
\leq 6$ GeV$^2$.
We show that the most relevant features 
arise already at LO. We will model the 
presence of NP using a non-zero 
contribution to the chirally flipped operator ${\cal O}_7^\prime $ 
according to the previous section.

Some important remarks concerning $A_T^2$ are in order here.
Eq.(\ref{eq:AT2LEET}) makes explicit several of the most important 
features of this observables, namely:

\begin{itemize}
\item
$A_T^2$ is sensitive to both the modulus and sign of $C^{{\rm eff} \prime}_{7}$, being approximately zero in the SM.
This sensitivity is enhanced by a factor $4 m_b M_B/q^2$ at low $q^2$ ($q^2\sim 1\, {\rm GeV}^2$), and for larger 
values of $q^2$ ($1<q^2<4 \,
 {\rm GeV}^2$) the observable decreases with a $1/s$ slope. This is clearly shown in Fig.1 looking at the 
curves, a,b,c and d.

\item Finally $A_T^2$ exhibits a zero, at the point $\Delta_{-}+\Delta_{+}^*=0$ corresponding exactly 
to the zero of the FB asymmetry at LO. Being this zero independent of $C^{{\rm eff} \prime}_7$,
all curves with SM-like $C_7$ should exhibit it (see Fig.1). 
Finally it was shown in \cite{ehmrr1} that contrary to the case of $A_T^2$ the
observable $A_{FB}$ does not show any remarkable sensitivity to the 
presence of RH currents.
This stress the importance of $A_T^2$ as one of the best indicators of the presence of this type of NP.

\end{itemize}

In summary $A_T^2$ provides different  informations depending on the region of $q^2$ 
analyzed: at low $q^2$ ($q^2\sim 1 \,{\rm GeV}^2$) basically sets the size of the coefficient $C^{{\rm eff}\prime}_{7}$ and 
at high $q^2$ ($q^2\sim 4 \rm \,GeV^2$) behaves as the FB asymmetry, with a zero in the 
energy axis. This last point implies obviously, that in case of a flipped sign solution for $C_7^{\rm 
eff}$ the behaviour of $A_T^2$ will change drastically. This is shown by 
the grey curve in Fig.1a that does not have 
a zero, like in  the FB asymmetry. In this sense $A_T^2$ goes beyond the $A_{FB}$ because it contains the most 
important features of this observable and also show up a dramatic dependence on the presence 
of RH currents ($O_7^\prime$)  invisible to $A_{FB}$.

\begin{figure}
\includegraphics[width=4cm,height=3cm]{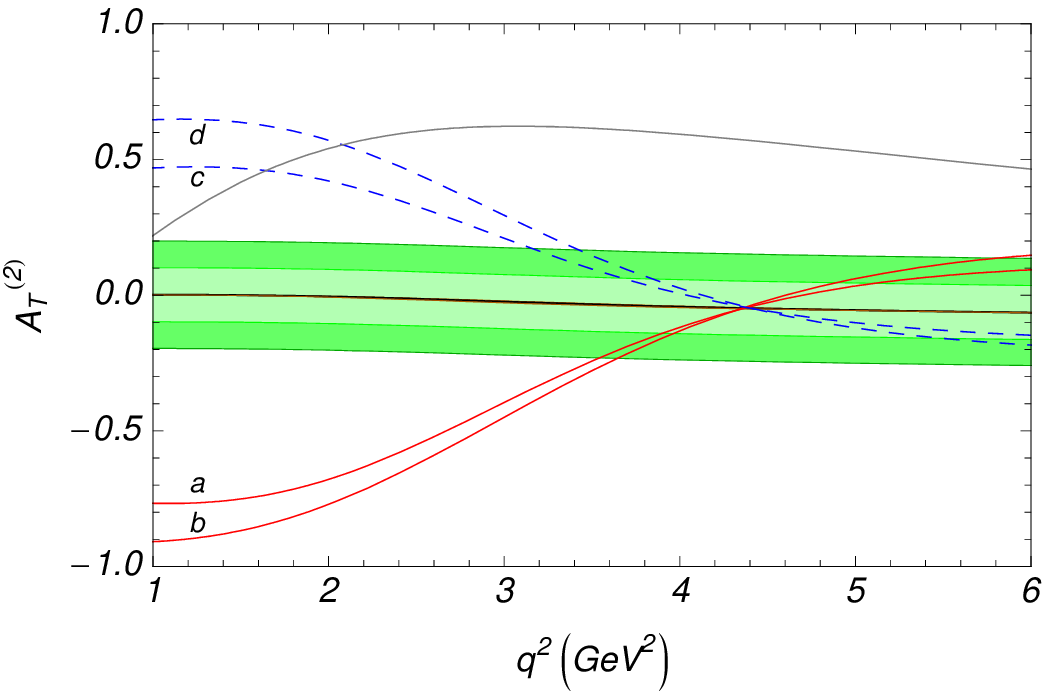}
\includegraphics[width=4cm,height=3cm]{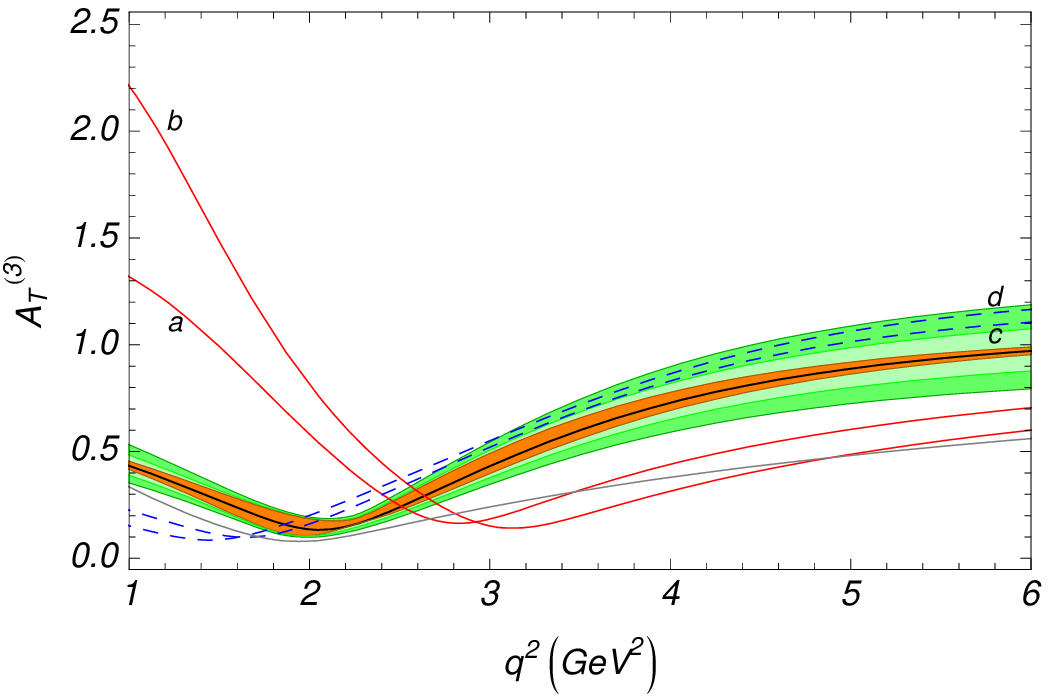}
\includegraphics[width=4cm,height=3cm]{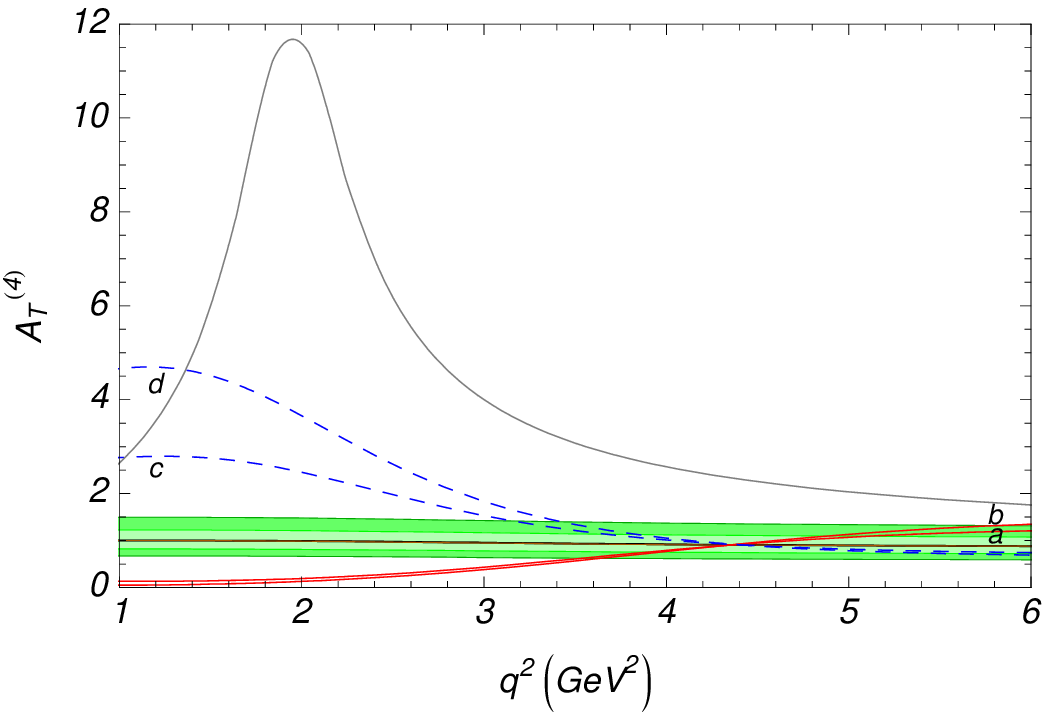}
\caption{$A_T^2$ (a), $A_T^3$(b) and $A_T^4$(c) in SM and SUSY
(curves 
\textit{a},\textit{b},\textit{c},\textit{d}). The outer dark and light (green) bands are, respectively, the 
possible 5 and 10\% $\Lambda/m_b$ corrections to the amplitudes, varied independently for each amplitude and 
added in quadrature. The inner (orange) band around the SM prediction (black curve) contains the hadronic and 
renormalisation scale uncertainties, also added in quadrature. The grey curve corresponds to the flipped sign 
solution $(C^{\rm{eff}}_7,C^{\rm{eff} \prime}_7)=(0.04,0.31).$} \end{figure}
A similar exercise can be done with the  observables $A_T^3$ and $A_T^4$ 
\cite{ehmrr1}. Those are 
particularly interesting because they open the sensitivity to the longitudinal spin amplitude $A_0$  minimizing, at the same time, 
the sensitivity to the other soft form factor $\xi_\|(0)$.

Both can be  very easily measured from the angular distribution and their explicit form in the heavy quark and large-$E_{K^*}$ 
limit for the spin amplitudes, even if less iluminating, still shows clearly the different way they 
depend on 
the RH currents. They are:
\begin{eqnarray}
   A_T^{{3}}\!=\!\frac{|A_{0L} A_{\| L}^* + A_{0R}^* A_{\| R}|}{\sqrt{|A_0|^2 |A_\perp|^2}}
&\!\!\sim\!\!&
\frac{ q^2 \left({\tilde \Delta^-}+f_1 \right) + {\tilde \Delta^-} \left(f_2+ {\tilde \Delta^-} f_3 
\right)}
{\sqrt{f_4 {\tilde \Delta}^{-2} + f_5 {\tilde \Delta}^- + f_6} \sqrt{{\tilde \Delta}^{+} q^2 + f_7 q^4 + 
f_8 {\tilde \Delta}^{+2}}}
  \, \nonumber \\
    A_T^{{4}}\!=\!\frac{ |A_{0L} A_{\perp L}^* - A_{0R}^* A_{\perp R}|   }{|A_{0L} A_{\| L}^* + A_{0R}^* 
A_{\| R}|}
&\!\!\sim\!\!&
\frac{ f_9 {\tilde \Delta}^+ + q^2 \left( f_{10} {\tilde \Delta}^- +f_{11}\right)      }{
 q^2 \left({\tilde \Delta^-}+f_1 \right) + {\tilde \Delta^-} \left(f_2+ {\tilde \Delta^-} f_3 \right)}
  \, \nonumber
  \end{eqnarray}
where $f_i=f({\cal C}_9^{\rm eff},C_{10})$ with $i=1,..11$ \cite{ehmrr2} are  simple functions of these coefficients  and ${{\tilde \Delta}^{\pm}={  
C^{{\rm eff}}_{7}}
 \pm { C^{{\rm eff} \prime}_{7}}}$. Here a small weak phase has been neglected.
As can be seen from their defining expressions and from  the Figs. 1b-1c, 
$A_T^3$ and $A_T^4$
play  a
complementary role, where $A_T^3$ shows a minimum $A_T^4$ has a maximum, and viceversa. For example, in the 
specific SUSY cases discussed, it is clear
from Figs. 1a-1b that the low-gluino scenario with negative mass insertion 
is clearly enhanced
in the
low-$q^2$ region for $A_T^4$, while the large-gluino scenario with positive mass insertion is clearly
signaled in the low-$q^2$ region for $A_T^3$ and in the large-$q^2$ region for $A_T^4$.
Finally, from the set of  functions $f_i$\cite{ehmrr2} it is easy at LO to
obtain with good precision  the position of the maxima/minima.
The way the sensitivity to RH currents is manifested is through the position of the maxima/minima 
in $A_T^3$ and $A_T^4$.
In the case of the flipped sign solution (grey curve of Fig. 1) this is 
shown also by the position of more prominent peak  in 
$A_T^4$ (as seen in Fig 1c).

The same philosophy as in \cite{ehmrr1} can be applied to CP violating 
observables (see 
\cite{ehmrr2,hurth} for a detailed discussion). Finally the excellent 
experimental sensitivity at LHCb specially for $A_T^2$ from a full angular 
analysis will allow to disentangle cleanly the presence of RH currents 
\cite{ehmrr1,ehmrr2,will}.\\

{\bf Acknowledgments:} JM and MR acknowledge financial support from FPA2008-01430 and MRTN-CT-2006-035482 and also
from UAB, UE and WR from STFC, TH acknowledges financial support from Heptools and the ITP
of the University Zurich.

\end{document}